\documentclass[amsmath,amssymb,superscriptaddress]{nature}
\usepackage{graphicx}
\usepackage[version=4]{mhchem}
\usepackage{microtype}
\usepackage{color}
\usepackage{caption}
\usepackage{bm}

\makeatletter
\let\saved@includegraphics\includegraphics
\AtBeginDocument{\let\includegraphics\saved@includegraphics}
\renewenvironment*{figure}{\@float{figure}}{\end@float}
\makeatother

\DeclareCaptionLabelSeparator{tate}{ $\mathbf{|}$ }
\title{Imaging the coupling between itinerant electrons and\\
  localised moments in the centrosymmetric skyrmion magnet \ce{GdRu2Si2}}

\author{Yuuki~Yasui$^{1,*}$, Christopher~J.~Butler$^{1, *}$,
  Nguyen~Duy~Khanh$^{1, 8}$, Satoru~Hayami$^{2, 3}$, Takuya~Nomoto$^2$,
  Tetsuo~Hanaguri$^{1, *}$, Yukitoshi~Motome$^2$, Ryotaro~Arita$^{1, 2}$,
  Taka-hisa~Arima$^{1, 4}$, Yoshinori~Tokura$^{1, 2, 5}$, \&
  Shinichiro~Seki$^{2, 6, 7}$}

\begin{document}

\maketitle

\begin{affiliations}
\item RIKEN Center for Emergent Matter Science, Wako, Saitama 351-0198, Japan
\item Department of Applied Physics, The University of Tokyo, Bunkyo, Tokyo 113-8656, Japan
\item Department of Physics, Hokkaido University, Sapporo, Hokkaido 060-0810, Japan
\item Department of Advanced Materials Science, The University of Tokyo, Kashiwa, Chiba 277-8561, Japan
\item Tokyo College, The University of Tokyo, Bunkyo-ku, Tokyo 113-8656, Japan
\item PRESTO, Japan Science and Technology Agency (JST), Kawaguchi 332-0012, Japan
\item Institute of Engineering Innovation, The University of Tokyo, Bunkyo, Tokyo 113-8656, Japan
\item Present address: Institute for Solid State Physics, The University of Tokyo, Chiba, 277-8581, Japan
\end{affiliations}

\noindent
\date{\today}

\newpage
\section*{Abstract}
\begin{abstract}
  Magnetic skyrmions were thought to be stabilised only in inversion-symmetry
  breaking structures, but skyrmion lattices were recently discovered in inversion
  symmetric Gd-based compounds, spurring questions of the stabilisation mechanism.
  A natural consequence of a recent theoretical proposal, a coupling between
  itinerant electrons and localised magnetic moments, is that the skyrmions are
  amenable to detection using even non-magnetic probes such as
  spectroscopic-imaging scanning tunnelling microscopy (SI-STM).
  Here SI-STM observations of \ce{GdRu2Si2} reveal patterns in the local density
  of states that indeed vary with the underlying magnetic structures.
  These patterns are qualitatively reproduced by model calculations which assume
  exchange coupling between itinerant electrons and localised moments.
  These findings provide a clue to understand the skyrmion formation mechanism
  in \ce{GdRu2Si2}.
\end{abstract}

\section*{Introduction}
Magnetic skyrmions are topologically-protected swirling spin structures
which have been observed in inversion-symmetry breaking structures, in which they
are stabilised by the Dzyaloshinskii-Moriya interaction
(DMI)\cite{muhlbauer_2009_science, yu_2010_nature, heinze_2011_np,
  seki_2012_science, nagaosa_2013_nn, kanazawa_2017_am}.
Recently discovered skyrmion lattices in inversion symmetric
crystals\cite{kurumaji_2019_sceence, hirschberger_2019_nc, khanh_2020_nn} have
been proposed to be stabilised instead by geometrical
frustration\cite{okubo_2012_prl, leonov_2015_nc},
or by multiple-spin interactions involving itinerant electrons\cite{martin_2008_prl,
  ozawa_2017_prl, hayami_2017_prb, takagi_2018_sa, hayami_2019_prb}.
The latter mechanism can be expected to apply in \ce{GdRu2Si2},
as it has a tetragonal structure belonging to the space group
\textit{I4/mmm}~[Fig.~\ref{fig1}(a)], in which geometrical frustration should be
absent.
The multiple-spin interactions have been theoretically argued to be mediated by
itinerant electrons\cite{ozawa_2017_prl, hayami_2017_prb}, but experimental
support is so far lacking.

\ce{GdRu2Si2} hosts a variety of magnetic orders\cite{garnier_1995_jmmm,
  garnier_1996_pb, samanta_2008_aipa}, whose localised  moments are provided by
Gd 4$f^7$ orbitals, and its itinerant electrons mostly come from Ru 4\textit{d}
orbitals with minor contributions from Si 3\textit{p} and Gd 5\textit{d}
orbitals\cite{nomoto_2020_prl}.
Very recently, resonant X-ray scattering (RXS) and Lorentz transmission electron
microscopy experiments have revealed the details of the magnetic structures of
the Gd moments, including the square skyrmion lattice, under magnetic field
applied parallel to the $c$-axis\cite{khanh_2020_nn}~[Fig.~\ref{fig1}(b)].
In this compound, the magnetic modulation vector $\mathbf{Q}_\mathrm{mag} =
(0.22, 0, 0)$ is observed to be common to all magnetic phases.
At low magnetic field (\mbox{Phase\,I}), a screw-like spin texture is realised.
In a narrow range between 2.1~T and 2.6~T (\mbox{Phase\,I\hspace{-1pt}I}),
the double-$Q$ square skyrmion lattice is stabilised, where the
magnetic structure can be approximately described by the superposition of two
screw spin structures with orthogonally arranged magnetic modulation vectors.
At higher magnetic field (\mbox{Phase\,I\hspace{-1pt}I\hspace{-1pt}I}),
a fan structure has been proposed while it has not been concluded whether this
\mbox{Phase\,I\hspace{-1pt}I\hspace{-1pt}I} is a single-$Q$ or double-$Q$ state.
Magnetic moments are fully polarised (FP) above around 10~T (FP phase).

When itinerant electrons are involved in the formation of the magnetic orders,
the relevant coupling between the itinerant electrons and localised magnetic
moments may enable the detection of information about the magnetic structure
through the charge channel.
To experimentally verify and gain insight into such coupling, we performed
spectroscopic-imaging scanning tunnelling microscopy (SI-STM) measurements on
\ce{GdRu2Si2}.
These revealed that the local density of states (LDOS) forms characteristic
spatial patterns that vary in accordance with magnetic structures, evidencing
the intimate coupling between itinerant electrons and localised magnetic moments.
The observed LDOS patterns clarify that not only \mbox{Phase\,I\hspace{-1pt}I}
but also \mbox{Phase\,I\hspace{-1pt}I\hspace{-1pt}I} hosts a double-$Q$ structure.
These patterns are reasonably reproduced by a model calculation which assumes
exchange coupling between itinerant electrons and localised magnetic moments.

\section*{Results}
\begin{figure}
  \centering
  \includegraphics[width=0.45\columnwidth]{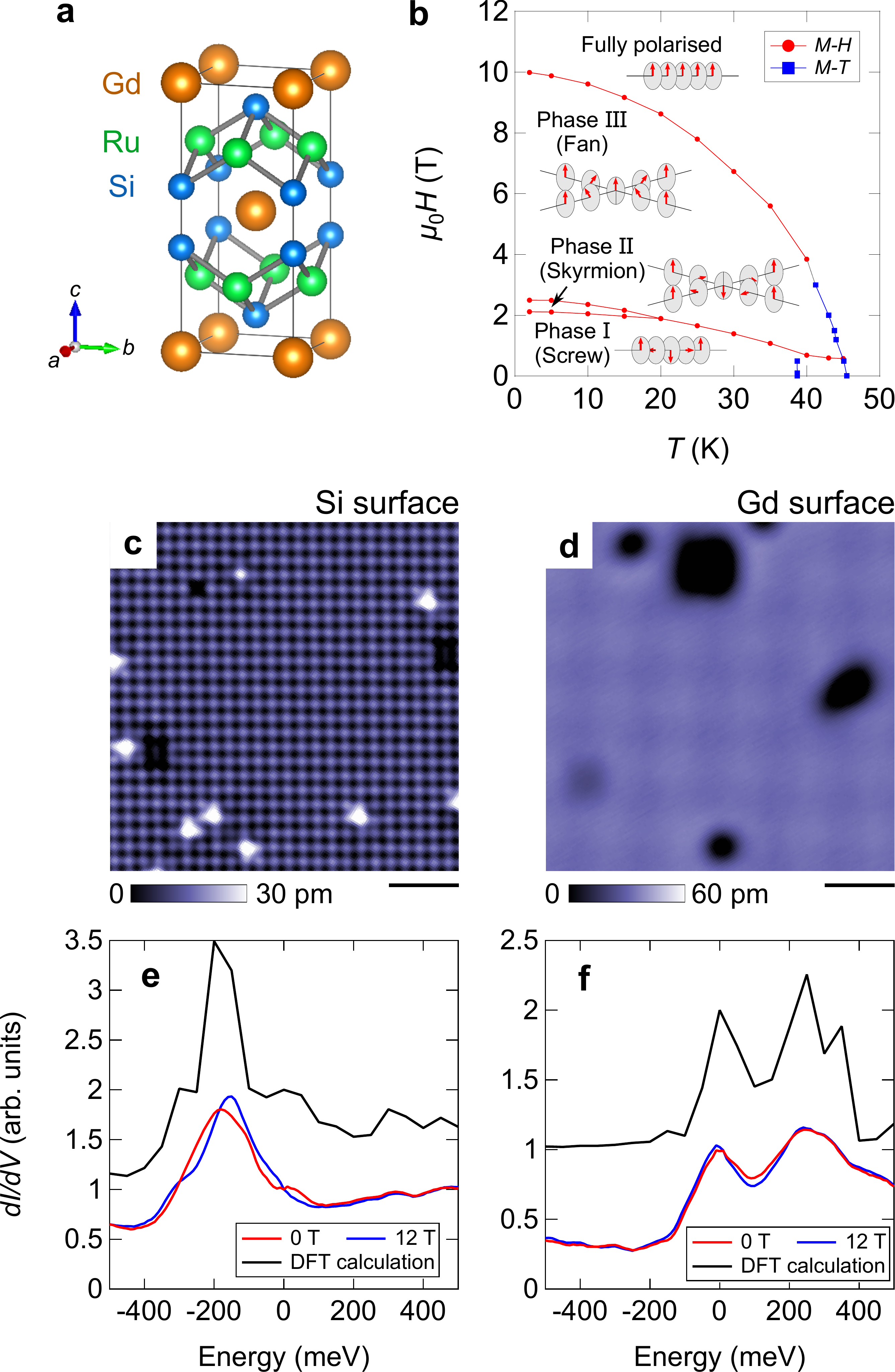}
  \caption{
    \textbf{Basic properties of \ce{GdRu2Si2}.}
    \textbf{a},~Crystal structure of \ce{GdRu2Si2}.
    \textbf{b},~Magnetic phase diagram for magnetic field $H$ applied parallel to the $c$-axis.
    Data points shown with red circles and blue squares were obtained using $H$ and
    $T$ dependence of the magnetisation, respectively\cite{khanh_2020_nn}.
    Schematics of the magnetic structures are depicted in insets.
    \textbf{c} and \textbf{d},~Constant-current topographic images of Si-terminated and
    Gd-terminated surfaces, respectively, at $T=1.5$~K and $\mu_0H=0$~T.
    The setup sample bias voltage was $V_{\mathrm{s}}=100$~mV, and tunnelling
    current was $I_{\mathrm{s}}=100$~pA. Scale bar 2 nm.
    \textbf{e} and \textbf{f},~Spatially averaged conductance spectra for Si-terminated and
    Gd-terminated surfaces, respectively.
    The setup condition is $V_{\mathrm{s}} = 500$~mV and $I_{\mathrm{s}} = 100$~pA.
    The bias modulation amplitude was $V_{\mathrm{mod}}=10$~mV.
    Calculated LDOS spectra for collinear ferromagnetic order at a tip position
    5~\AA \ above the surface are also shown and vertically shifted for clarity.
  }
  \label{fig1}
\end{figure}

We inspected multiple cleaved surfaces and observed two types of termination as
shown in Figs.~\ref{fig1}(c) and (d).
One of the terminations showed a clear atomic lattice with the lattice constant
corresponding to either Gd-Gd or Si-Si in-plane distance [Fig.~\ref{fig1}(c)].
This suggested that cleavage occured between Gd and Si layers.
Atomic corrugations were hardly seen on the other termination surface even
with the identical scanning tip [Fig.~\ref{fig1}(d)].
Among seven samples we investigated, we did not observe any surface with atomic
corrugations corresponding to Ru-Ru lattice spacing.

Figures~\ref{fig1}(e) and (f) show that two types of surfaces exhibit different
tunnelling conductance $dI/dV$ spectra, which are taken at surfaces shown in
Figs.~\ref{fig1}(c) and (d), respectively.
Here, $I$ is tunnelling current, and $V$ is sample bias voltage.
To identify the termination, the observed $dI/dV$ spectra, which are
proportional to the LDOS at given tip height, are compared with first-principle calculations.
The calculations are performed based on the density functional theory (DFT) for
slab systems, where we assume collinear ferromagnetic order.
The overall correspondence allows us to assign the termination of
Fig.~\ref{fig1}(c) to Si, and that of Fig.~\ref{fig1}(d) to Gd.
Hereafter, we will discuss the Si-terminated surface since the atomic and
electronic modulations are more clearly observed for this surface.
Additionally, the Gd-terminated surface shows properties different
from those expected from the bulk behaviour, which may be induced by surface
effects (see Supplementary Note 1).

\begin{figure*}
  \centering
  \includegraphics[width=0.8\columnwidth]{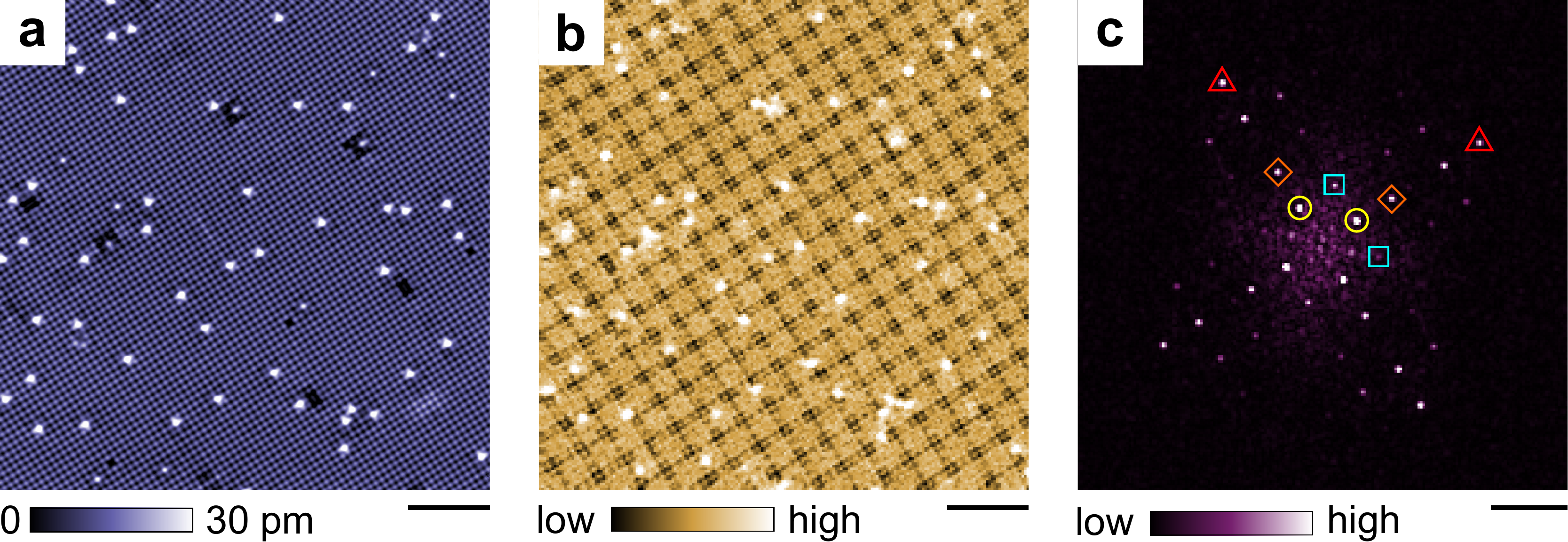}
  \caption{
    \textbf{SI-STM results on the Si-terminated surface at $\bm{T=1.5}$~K and
      $\bm{\mu_0H = 2.3}$~T in \mbox{Phase\,I\hspace{-1pt}I} (skyrmion)}.
    \textbf{a},~Constant-current topographic image.
    The setup condition is $V_{\mathrm{s}} = 100$~mV and $I_{\mathrm{s}} =
    100$~pA.
    Scale bar 5 nm.
    \textbf{b},~Normalised conductance map $L(\mathbf{r}, E = -20~\mathrm{meV})$ taken in the same
    field of view.
    $V_{\mathrm{s}} = 100$~mV, $I_{\mathrm{s}} = 100$~pA, and
    $V_{\mathrm{mod}}=5$~mV.
    Scale bar 5 nm.
    \textbf{c},~Fourier transform (FT) of (b).
    Peaks appear at $\mathbf{Q}_1=(0.22, 0)$ and $\mathbf{Q}_2=(0, 0.22)$ (yellow circles),
    $\mathbf{Q}_1 \pm \mathbf{Q}_2$ (blue squares), $2\mathbf{Q}_1$ and $2\mathbf{Q}_2$ (orange
    diamonds), and atomic Bragg points $\mathbf{G} \equiv (1,0)$ and $(0,1)$
    (red triangles).
    Scale bar 1 nm$^{-1}$.
  }
  \label{fig2}
\end{figure*}

To investigate the impact of the magnetic order on the charge channel, SI-STM is
performed in a magnetic field range that covers all of the magnetic phases at
low temperature.
For the spectroscopic imaging, full $I(V)$ and
$\frac{dI(V)}{dV}$ curves were recorded at each pixel to obtain
spatial dependence and bias dependence simultaneously.
We analyse normalised conductance maps
$L(\mathbf{r},E=eV) \equiv \frac{dI(\mathbf{r},V)}{dV} / \frac{I(\mathbf{r},V)}{V}$
instead of raw conductance maps $\frac{dI(\mathbf{r},V)}{dV}$ to suppress
artifacts from the constant-current feedback loop\cite{feenstra_1987_ss, kohsaka_2007_science}.
Here, $\mathbf{r}$ is lateral position, and $e$ is the elementary charge.
We begin our discussion from \mbox{Phase\,I\hspace{-1pt}I},
which is identified as the square skyrmion lattice phase.
Figures~\ref{fig2}(a) and (b) show a constant-current topograph and
a $L(\mathbf{r},E=-20~\mathrm{meV})$ map in the same field of view, respectively.
In the $L(\mathbf{r},E)$ map, a four-fold symmetric superstructure with a period of
1.9~nm is observed.
This period corresponds to that of the skyrmion lattice previously determined
using RXS\cite{khanh_2020_nn}.
Therefore, we infer that the pattern of the skyrmion lattice is
imprinted in the LDOS of itinerant electrons.
As shown in Fig.~\ref{fig2}(c), Fourier analysis clarifies periodic
components in the $L(\mathbf{r}, E)$ map.
In addition to atomic Bragg peaks at $\mathbf{G} \equiv (1, 0)$ and $(0, 1)$, several
modulation vectors $\mathbf{Q}$ are observed.
Modulations with the smallest $|\mathbf{Q}|$ are found at $\mathbf{Q}_1=(0.22, 0)$ and
$\mathbf{Q}_2=(0, 0.22)$.
We also observed peaks at $\mathbf{Q}_1 \pm \mathbf{Q}_2$, $2\mathbf{Q}_1$, and
$2\mathbf{Q}_2$.
Other peaks are assigned to `replicas' of these \textit{Q}-vectors shifted by $\mathbf{G}$.
(see Supplementary Figure~3 for higher spatial resolution data,
Supplementary Figure~4 for bias voltage dependence,
and Supplementary Figure~5 for location dependence of the spectra.)

LDOS patterns in the other magnetic phases are also investigated.
The electronic modulations clearly change depending on the magnetic phase,
as seen in $L(\mathbf{r},E=-20~\mathrm{meV})$ maps [Figs.~\ref{fig3}(a)-(d)] and
their Fourier transformed images [Figs.~\ref{fig3}(e)-(h)].
(see Supplementary Figure~6 for raw $dI/dV$ maps at different magnetic fields
and Supplementary Figure~7 for the data indicating the robustness of the tip
throughout the measurements.)
In \mbox{Phase\,I}, the LDOS forms a two-fold symmetric pattern, which is
composed of modulation vectors $2\mathbf{Q}_1$ and $\mathbf{Q}_1+\mathbf{Q}_2$.
In \mbox{Phase\,I\hspace{-1pt}I\hspace{-1pt}I}, the LDOS pattern is four-fold symmetric
and is characterised by $2\mathbf{Q}_1$ and $2\mathbf{Q}_2$.
While the previous spatial-averaging RXS experiments cannot distinguish a
double-$Q$ order and a multiple-domain state of single-$Q$ order, the present
real-space imaging clarifies a double-\textit{Q} order is realised in
\mbox{Phase\,I\hspace{-1pt}I\hspace{-1pt}I}.
In the FP phase, all $Q$-vectors disappear except for atomic Bragg peaks.
It should be noted that $\mathbf{Q}_1$ and $\mathbf{Q}_2$ modulations are
observed only in \mbox{Phase\,I\hspace{-1pt}I}.
We discuss this point below.

\begin{figure*}
  \centering
  \includegraphics[width=0.9\columnwidth]{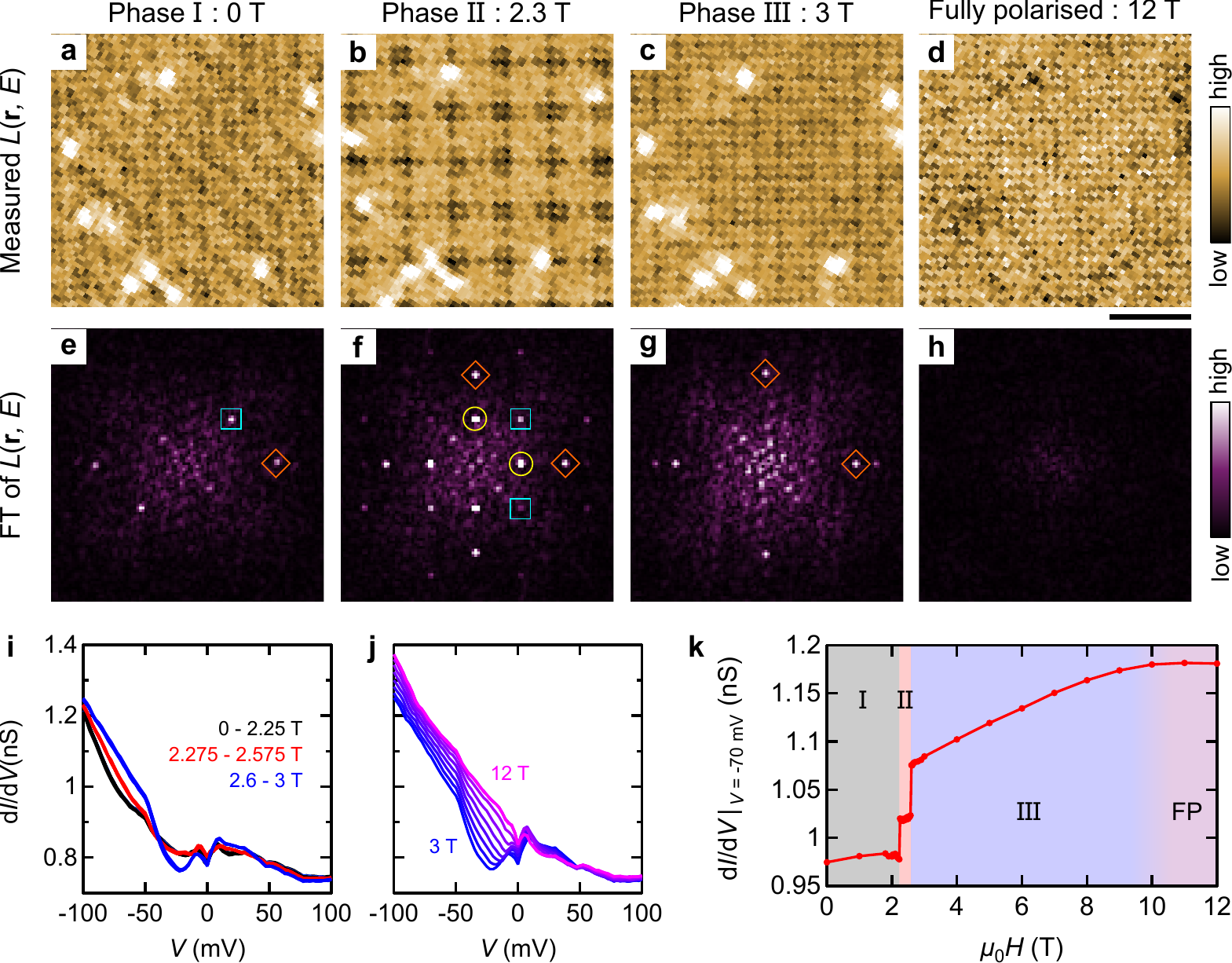}
  \caption{
    \textbf{SI-STM for different magnetic phases.}
    \textbf{a-d},~Normalised conductance maps $L(\mathbf{r},E = -20~\mathrm{meV})$ in
    different magnetic phases measured at 1.5~K.
    The images were cropped from $30 \times 30~\mathrm{nm^2}$-large maps.
    $V_{\mathrm{s}} = 100$~mV, $I_{\mathrm{s}} = 100$~pA, and
    $V_{\mathrm{mod}}=5$~mV.
    Scale bar 3 nm.
    \textbf{e-h},~FTs of $L(\mathbf{r},E = -20~\mathrm{meV})$ maps.
    \textbf{i} and \textbf{j},~Spatially averaged $dI/dV$ curves \textbf{i}~below 3~T
    and \textbf{j}~above 3~T.
    A spectrum was taken every 25~mT between 2 and 2.7~T.
    $V_{\mathrm{s}} = 100$~mV, $I_{\mathrm{s}} = 100$~pA, and $V_{\mathrm{mod}}
    =2.5$~mV.
    \textbf{k},~Magnetic field dependence of the value of $dI/dV$ at $V=-70$~mV.
  }
  \label{fig3}
\end{figure*}

To further corroborate the correspondence between the LDOS and magnetic orders,
we investigate detailed magnetic-field dependence of the $dI/dV$ spectrum.
Figure~\ref{fig3}(i) shows a series of spatially averaged $dI/dV$ spectra for
$\mu_0H \leq3$~T.
The spectrum varies only subtly within each magnetic phase.
On the other hand, it exhibits discontinuous changes across phase boundaries at
2.275~T and 2.6~T.
Note that these transition fields slightly change depending on the field history.
Such first-order-like transitions are consistent with the magnetic
measurement\cite{khanh_2020_nn}.
They also reflect the transition between topologically trivial and non-trivial phases.
Above 3~T, as shown in Fig.~\ref{fig3}(j), the spectrum evolves continuously
until it saturates in the FP phase above 10~T.
The trend is clearly seen by plotting magnetic field dependence of $dI/dV$ at a
chosen energy $E=-70$~meV [Fig.~\ref{fig3}(k)].
(see Supplementary Figure~8 for the $dI/dV$ evolution at different energy,
Supplementary Figure~9 for the data taken with decreasing field,
and Supplementary Figure~10 for the same analyses for the normalised conductance.)
The observed one-to-one correspondence between the LDOS and the magnetic phase
indicates that itinerant electrons and localised magnetic moments are intimately coupled.

Let us compare the periods of observed LDOS modulations with previously reported
magnetic structures\cite{khanh_2020_nn}.
In the LDOS maps, fundamental modulations of $\mathbf{Q}_1$ and $\mathbf{Q}_2$
appear only in \mbox{Phase\,I\hspace{-1pt}I} while $2\mathbf{Q}_1$ and/or
$2\mathbf{Q}_2$ show up in all the magnetic phases except for the FP phase.
Namely, the LDOS takes on the period of the magnetic structure in
\mbox{Phase\,I\hspace{-1pt}I};
the LDOS period becomes a half of the magnetic period in \mbox{Phase\,I} and
\mbox{I\hspace{-1pt}I\hspace{-1pt}I}.
One may expect halved charge period in systems with coupled charge- and spin-density waves,
where itinerant electrons host both spin and charge modulations\cite{zachar_1998_prb}.
However, such a simple relation in periods does not apply for
\mbox{Phase\,I\hspace{-1pt}I} (skyrmion lattice) of \ce{GdRu2Si2}.
The absence of $\mathbf{Q}_1$ modulation in \mbox{Phase\,I} ensures that
the scanning tip is not magnetized due to unintentional pick-up of magnetic Gd atoms.

\section*{Discussion}
\begin{figure*}
  \centering
  \includegraphics[width=0.9\columnwidth]{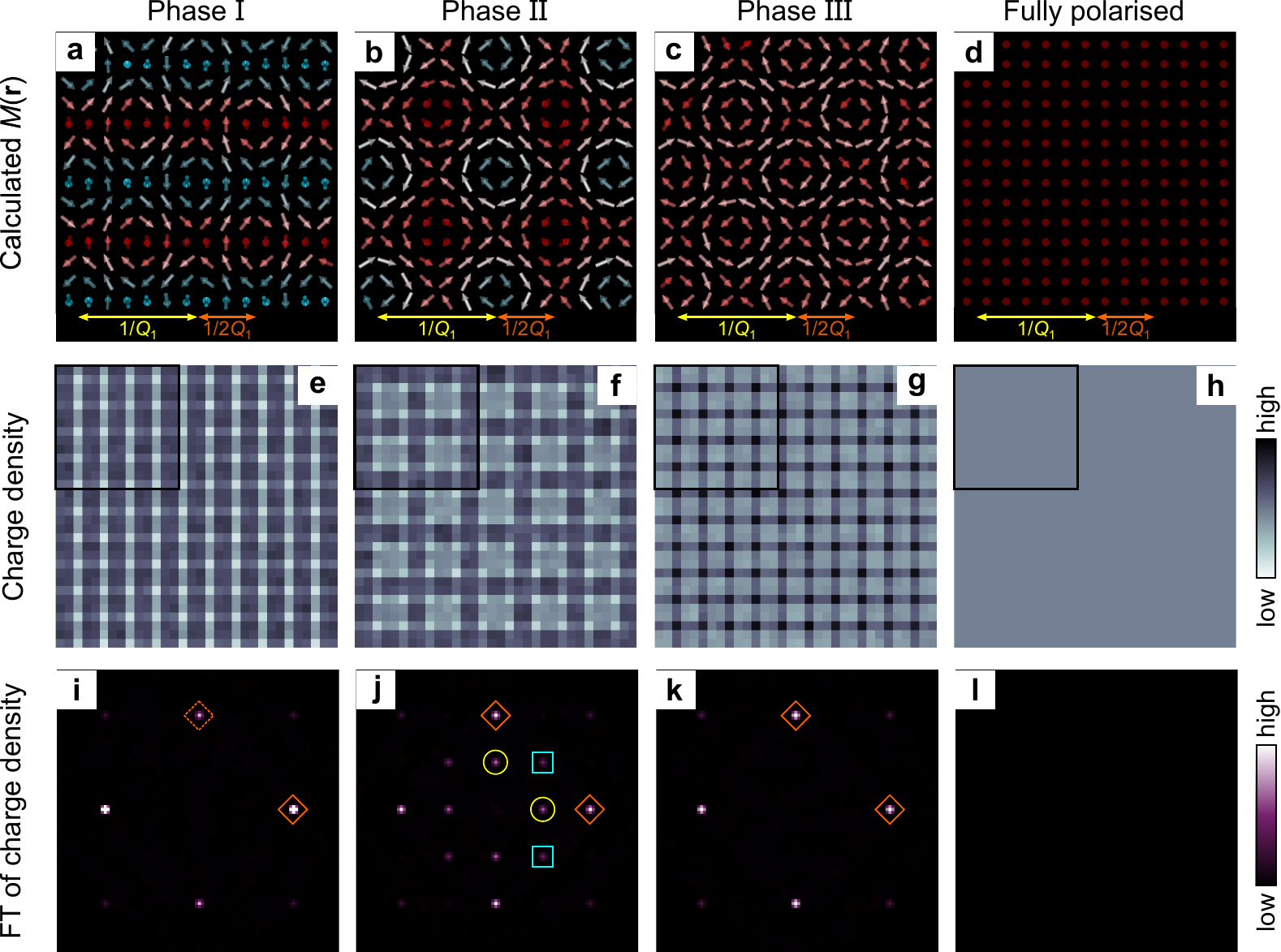}
  \caption{
    \textbf{Magnetic and electronic structure by the model calculation.}
    \textbf
    {a-d},~Calculated local magnetic moments $M(\mathbf{r})$ in different magnetic
    phases.
    Red (cyan) colour represents the magnetic moments pointing (anti)parallel to
    the magnetic field.
    \textbf{e-h},~Calculated charge-density patterns in different magnetic phases.
    The regions shown in (a)-(d) are marked with black squares.
    \textbf{i-l},~FTs of (e)-(h).
    Coloured symbols correspond to those in Fig.~\ref{fig2}.
  }
  \label{fig4}
\end{figure*}

In order to understand the origin of the observed LDOS modulations,
we performed calculations for magnetic configurations and charge-density
distributions.
The magnetic configurations are obtained for an effective spin model with
long-range interactions that can be originated from the coupling between the
itinerant electron spins and localised spins.
The Hamiltonian is given as\cite{hayami_2017_prb}
\begin{align}
  \label{eq:Model}
  \mathcal{H}=  2\sum_\nu
  \left[ -J  \left(\sum_{\alpha=x,y,z}\Gamma^{\alpha\alpha}_\mathbf{Q_{\nu}} S^\alpha_{\mathbf{Q_{\nu}}} S^\alpha_{-\mathbf{Q_{\nu}}}\right)
  +\frac{K}{N} \left(\sum_{\alpha=x,y,z}\Gamma^{\alpha\alpha}_\mathbf{Q_{\nu}} S^\alpha_{\mathbf{Q_{\nu}}} S^\alpha_{-\mathbf{Q_{\nu}}}\right)^2 \right]+H \sum_i S_i^z,
\end{align}
where $\mathbf{S}_{\mathbf{Q}_\nu} =(S^x_\mathbf{Q_{\nu}}, S^y_\mathbf{Q_{\nu}},
S^z_\mathbf{Q_{\nu}})$ is the Fourier transform of the localised spin
$\mathbf{S}_i$ treated as a classical vector with the normalisation
$|\mathbf{S}_i|=1$, and $N$ is the system size.
The Hamiltonian includes two exchange terms defined in momentum space:
the bilinear exchange interaction $J$ and the biquadratic exchange interaction $K$.
The wave numbers $\mathbf{Q}_\nu$ are set to be $\mathbf{Q}_1=(\pi/3,0)$ and
$\mathbf{Q}_2=(0,\pi/3)$.
We also introduce an anisotropy due to the symmetry of the tetragonal crystal
structure  as
$\Gamma^{yy}_{\mathbf{Q}_1}=\Gamma^{xx}_{\mathbf{Q}_2}=\gamma_1$,
$\Gamma^{xx}_{\mathbf{Q}_1}=\Gamma^{yy}_{\mathbf{Q}_2}=\gamma
_2$, and
$\Gamma^{zz}_{\mathbf{Q}_1}=\Gamma^{zz}_{\mathbf{Q}_2}=\gamma_3$, which selects the
spiral plane.
The last term in Eq.~(\ref{eq:Model}) represents the Zeeman coupling to an
external magnetic field $H$.
Performing the simulated annealing by means of Monte Carlo simulations for the
$N=96^2$ sites at  $J=1$, $K=0.5$, $\gamma_1=0.9$, $\gamma_2=0.72$, and
$\gamma_3=1$, we obtained the screw, skyrmion lattice, fan, and fully polarised
states while increasing $H$.
We show the spin configurations for each phase at $H=0$,
$0.6$, $0.725$, and $\infty$ in Figs.~\ref{fig4}(a), \ref{fig4}(b),
\ref{fig4}(c), and \ref{fig4}(d), respectively.

The square skyrmion lattice (\mbox{Phase\,I\hspace{-1pt}I}) is found to be
stabilized with the help of the anisotropy for the tetragonal crystal structure.
Note that square skyrmion lattices in square crystal structures were not
expected to be stabilized in previous reports, in which the magnetic anisotropy
was not considered\cite{hayami_2017_prb}.
At a higher magnetic field, the calculation predicts a double-$Q$ fan structure
in \mbox{Phase\,I\hspace{-1pt}I\hspace{-1pt}I}, consistent with the experiments
(Figs.~\ref{fig3}c and \ref{fig3}g).

The charge density is calculated by considering itinerant electrons coupled with
the spin textures obtained as above.
The Hamiltonian is given as
\begin{align}
  \label{eq:Ham}
  \mathcal{H} =
  -t\sum_{\langle i, j \rangle,  \sigma}  (c^{\dagger}_{i\sigma}c_{j \sigma}+ \mathrm{h.c.})
  +J_\mathrm{K} \sum_{i} \mathbf{s}_i \cdot \mathbf{S}_i,
\end{align}
where $c^{\dagger}_{i\sigma}$ ($c_{i \sigma}$) is the creation (annihilation)
operator of an itinerant electron at site $i$ and with spin $\sigma$.
The first term represents the nearest-neighbour hopping of electrons.
The second term represents the spin-charge coupling between the electron spin
$\mathbf{s}_i=(1/2)\sum_{\sigma, \sigma'}c^{\dagger}_{i\sigma} \mathbf{\sigma}_{\sigma
  \sigma'} c_{i \sigma'}$ and the underlying spin texture;
$\mathbf{\sigma}$ denotes the Pauli matrix.
We set $t=J_\mathrm{K}=1$ and the chemical potential $\mu=-3$.
The charge density at site $i$, $\langle n_{i} \rangle =\langle \sum_\sigma
c^{\dagger}_{i\sigma}c_{i\sigma} \rangle$, is obtained by diagonalising the
Hamiltonian in Eq.~(\ref{eq:Ham}) for each spin texture.
The results and their Fourier transforms are shown in Figs.~\ref{fig4}(e)-(l).
(see Supplementary Figure~11 for the results with different chemical potentials.)

By comparing Figs.~\ref{fig4}(a)-(d) and (e)-(h), it can be seen
that charge-distribution patterns reflect the magnetic structures.
This can be interpreted as follows.
Since the itinerant electrons' spins are aligned with localised moments, kinetic
energy of the itinerant electrons depends on the relative angle between
localised magnetic moments at neighbouring sites.
Thus, itinerant electrons reflect local magnetic structures.
The charge modulations on the magnetic textures are qualitatively understood
from the scattering process via the spin-charge coupling $J_\mathrm{K}$.
Within the second-order perturbation theory, the charge density at momentum
$\mathbf{q}$ is proportional to $J_\mathrm{K}^2 \sum_{\mathbf{q}_1 \mathbf{q}_2}
\Lambda_{\mathbf{q}_1\mathbf{q}_2} (\mathbf{S}_{\mathbf{q}_1}\cdot
\mathbf{S}_{\mathbf{q}_2})\delta_{\mathbf{q},\mathbf{q}_1+\mathbf{q}_2}$, where
$\Lambda_{\mathbf{q}_1\mathbf{q}_2}$ is a form factor depending on the
electronic structure and $\delta$ is the Kronecker delta.
The nonzero $\mathbf{S}_{\mathbf{q}}$ components in each magnetic texture satisfying $
\mathbf{S}_{\mathbf{q}_1}\cdot \mathbf{S}_{\mathbf{q}_2} \neq 0$ explain the wave numbers
$\mathbf{q}$ for the charge modulations.

The calculated charge modulations resemble the basic features of the observed LDOS
structures.
The wavy modulation orthogonal to the screw structure in \mbox{Phase\,I}
results in the stripe pattern in charge density [Fig.~\ref{fig4}(e)].
$2\mathbf{Q}_1$ and $2\mathbf{Q}_2$ appear in all the magnetic phases except for
the FP phase and dominate in \mbox{Phase\,I} and \mbox{I\hspace{-1pt}I\hspace{-1pt}I}
[Figs.~\ref{fig4}(i)-(l)].
This is because the local configuration of relative angles between neighbouring
spins becomes almost the same every half periodicity of the magnetic modulations.
In contrast, in \mbox{Phase\,I\hspace{-1pt}I}, the angles between
neighbouring spins at the skyrmion core and in between the cores are different.
Therefore, $\mathbf{Q}_1$ and $\mathbf{Q}_2$ modulations appear in the charge sector.
It should be noted that the peak at $\mathbf{Q}_1+\mathbf{Q}_2$ in \mbox{Phase\,I}
cannot be explained by the present model, and more advanced model may be
necessary to explain this behaviour.
Nevertheless, the overall good agreement between the observed and calculated
spatial patterns in the double-$Q$ states suggests that the present theoretical
framework based on multiple-spin interactions  well captures the physics behind
the skyrmion formation in this centrosymmetric magnet.

We note that magnetic structures, including skyrmions, have also been detected
with non-magnetic STM tips via the mechanisms known as the tunnelling anisotropic
magnetoresistance (TAMR)\cite{bode_2002_prl, heinze_2011_np,
  vonBergmann_2012_prb,vonBergmann_2015_nl, hanneken_2015_nn} and the non-colinear
magnetoresistance (NCMR)\cite{hanneken_2015_nn, kubetzka_2017_prb} in 3\textit{d}
transition metals where the magnetic structures are originated from magnetic
moments carried by itinerant electrons.
The TAMR effect may not explain the present observation.
This is because the centre and the edges of skyrmions show different contrast
in the present LDOS map, whereas spins pointing in and out of the surface should
appear similarly for TAMR effect.
By contrast, the observed LDOS modulations are similar to those caused by the
NCMR.
In the case of \ce{GdRu2Si2}, however, NCMR effect alone is not enough to explain the
present observation because the coupling between itinerant electrons and
localised moments is indispensable.
Our observations evidence such a coupling, which not only allows us to access
the localised moments from the charge sector but also may play a role for the
itinerant-electron mediated magnetic interactions responsible for the skyrmion
formation.

In conclusion, our observation of modulations of itinerant electrons associated
with magnetic structures provides evidence for a coupling
between itinerant electron states and local magnetic moments
in the centrosymmetric skyrmion magnet \ce{GdRu2Si2}.
The observed modulations are reproduced by charge density calculations
which consider exchange coupling between itinerant electrons
and localised magnetic moments fixed by anisotropic multiple-spin interactions.
We interpret that this happens because spatially varying kinetic energy of
itinerant electrons reflects neighbouring configurations of Gd moments.
These results together have established the basic framework of the coupling
between itinerant electrons and local magnetic moments in \ce{GdRu2Si2}.
Further theoretical and experimental investigation is required to explain the
detailed features in the observed modulations (such as
$\mathbf{Q}_1+\mathbf{Q}_2$ component in Phase\,I), which may also lead to
identify the microscopic formation mechanism of the square skyrmion lattice in
the absence of the DM interaction.

\begin{methods}
  \subsection{Sample preparation and STM measurements}
  \ce{GdRu2Si2} single crystals were grown with the floating zone method\cite{khanh_2020_nn}.
  The samples were cleaved in an ultra-high vacuum chamber ($\sim 10^{-10}$~Torr)
  at around 77~K to expose clean and flat (001) surfaces and then transferred to
  the microscope\cite{hanaguri_2006_jpcs} without breaking vacuum.
  As scanning tips, tungsten wires were used after electro-chemical etching in
  \ce{KOH} aqueous solution, followed by tuning using field ion microscopy and
  controlled indentation at clean Cu(111) surfaces.
  All the measurements were conducted at temperature $T\simeq 1.5$~K, and magnetic field
  was applied along the crystalline \textit{c}-axis.
  Tunnelling conductance was measured using the standard lock-in technique with AC
  frequency of 617.3~Hz.

  \subsection{Calculation of the density of states}
  The local density of states shown in the main text are obtained from first
  principles calculations for slab systems. The actual calculations are performed
  based on density functional theory (DFT) with VASP code\cite{kresse_1993_prb,
    kresse_1994_prb}, where we assume a collinear ferromagnetic order.
  We consider the conventional cell of \ce{GdRu2Si2} with the experimental lattice
  parameters\cite{hiebl_1983_jmmm}, $a=4.1634$~\AA, $c=9.6102$~\AA, and
  $z_{\mathrm{Si}}=0.375$, and then, stack it to construct the supercell systems
  with eight Ru-layers.
  Finally, we insert a vacuum layer with 10~\AA \ at the edge of the slabs, and
  perform a surface relaxation calculation to optimize the positions of surface
  atoms.
  The LDOS spectra are calculated as the
  summation of partial charge densities of the Bloch states,
  $\sum_{n\mathbf{k}}'|\psi_{n\mathbf{k}}(\mathbf{r})|^2$, where the summation
  $\sum_{n\mathbf{k}}'$ is restricted to $(n\mathbf{k})$ with the energy
  $\varepsilon_{n\mathbf{k}} \in [\varepsilon-\Delta, \varepsilon+\Delta]$.
  We employ the exchange-correlation functional proposed by
  Perdew \textit{et al.}\cite{perdew_1996_prl}, $E_\mathrm{c}=450$~eV as the
  cutoff energy for the planewave basis set, and
  $N_\mathbf{k}=10\times10\times1$ as the number of $\mathbf{k}$-points for the
  self-consistent calculation.
  In the LDOS calculations, we use a denser k-mesh,
  $N_\mathbf{k}=40\times40\times1$ and $\Delta=25$ meV.

\end{methods}

\begin{addendum}
\item The authors acknowledge M. Hirschberger, K. Ishizaka, Y. Kohsaka, T.
  Machida, and Y. Ohigashi for discussion.
  This work was supported by JST CREST Grant Nos.~JPMJCR16F2
  , JPMJCR18T2
  , and JPMJCR1874
  , by Grant-in-Aid JSPS KAKENHI Grant Nos.~JP19H05824
  , JP19H05825
  , JP19H05826
  , JP18K13488
  , JP20H00349
  , and JP18H03685
  , by JST PRESTO Grant No.~JPMJPR18L5
  , and by Asahi Glass Foundation.
  C.J.B. acknowledges support from RIKEN's SPDR fellowship.

\item[Data availability] The data that support the findings of this study are
  available from the corresponding author upon reasonable request.

\item[Code availability] The codes used for this study are available from the
  corresponding author upon reasonable request. 

\item[Contributions] T.~H., T.-h.~A., Y.~T., and  S.~S. conceived the project.
  N.~D.~K. synthesised \ce{GdRu2Si2} single crystals.
  Y.~Y., C.~J.~B., and T.~H. carried out STM measurements and analysed the
  experimental data.
  S.~H. and Y.~M. carried out model calculations.
  T.~N. and R.~A. carried out LDOS calculations.
  Y.~Y, C.~J.~B., S.~H., T.~N., T.~H., and S.~S. wrote the manuscript with inputs
  from all the authors.

\item[Corresponding author] Correspondence to Y. Y.~(email: yuuki.yasui@riken.jp), C.
  J. B~(email: christopher.butler@riken.jp), and T. H.~(email: hanaguri@riken.jp)

\item[Competing interests] The authors declare no competing interests.

\item[Supplementary information] SI-STM data for Gd-terminated surface and
  additional data supporting the main observation are presented.
\end{addendum}

\end{document}


\title{\textmd{Supplementary Information for}\\
  Imaging the coupling between itinerant electrons and localised moments\\in the centrosymmetric skyrmion magnet \ce{GdRu2Si2}}

\author{Yuuki~Yasui}
\affiliation{RIKEN Center for Emergent Matter Science, Wako, Saitama 351-0198, Japan}
\author{Christopher~J.~Butler}
\affiliation{RIKEN Center for Emergent Matter Science, Wako, Saitama 351-0198, Japan}
\author{Nguyen~Duy~Khanh}
\affiliation{RIKEN Center for Emergent Matter Science, Wako, Saitama 351-0198, Japan}
\author{Satoru~Hayami}
\affiliation{Department of Applied Physics, The University of Tokyo, Bunkyo, Tokyo 113-8656, Japan}
\affiliation{Department of Physics, Hokkaido University, Sapporo, Hokkaido 060-0810, Japan}
\author{Takuya~Nomoto}
\affiliation{Department of Applied Physics, The University of Tokyo, Bunkyo, Tokyo 113-8656, Japan}
\author{Tetsuo~Hanaguri}
\affiliation{RIKEN Center for Emergent Matter Science, Wako, Saitama 351-0198, Japan}
\author{Yukitoshi~Motome}
\affiliation{Department of Applied Physics, The University of Tokyo, Bunkyo, Tokyo 113-8656, Japan}
\author{Ryotaro~Arita}
\affiliation{RIKEN Center for Emergent Matter Science, Wako, Saitama 351-0198, Japan}
\affiliation{Department of Applied Physics, The University of Tokyo, Bunkyo, Tokyo 113-8656, Japan}
\author{Taka-hisa~Arima}
\affiliation{RIKEN Center for Emergent Matter Science, Wako, Saitama 351-0198, Japan}
\affiliation{Department of Advanced Materials Science, The University of Tokyo, Kashiwa, Chiba 277-8561, Japan}
\author{Yoshinori~Tokura}
\affiliation{RIKEN Center for Emergent Matter Science, Wako, Saitama 351-0198, Japan}
\affiliation{Department of Applied Physics, The University of Tokyo, Bunkyo,
  Tokyo 113-8656, Japan}
\affiliation{Tokyo College, The University of Tokyo, Bunkyo-ku, Tokyo 113-8656, Japan}
\author{Shinichiro~Seki}
\affiliation{Department of Applied Physics, The University of Tokyo, Bunkyo, Tokyo 113-8656, Japan}
\affiliation{PRESTO, Japan Science and Technology Agency (JST), Kawaguchi 332-0012, Japan}
\affiliation{Institute of Engineering Innovation, The University of Tokyo, Bunkyo, Tokyo 113-8656, Japan}

\date{\today}

\maketitle

\section{S\lowercase{pectroscopic-imaging} STM \lowercase{on} G\lowercase{d-terminated surface}}

\begin{figure}[t]
  \includegraphics[width=0.9\columnwidth]{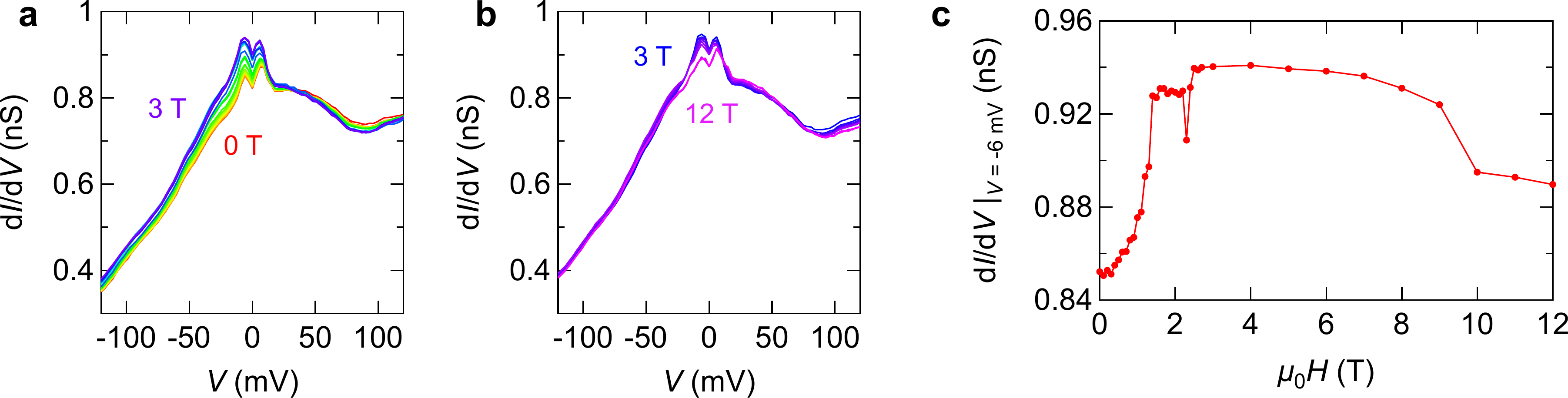}
  \caption{
    Spatially averaged $dI/dV$ curves of the Gd-terminated surface \textbf{a}~below
    3~T and \textbf{b}~above 3~T measured at 1.5~K.
    $V_{\mathrm{s}} = 120$~mV, $I_{\mathrm{s}} = 100$~pA, and $V_{\mathrm{mod}}
    =3$~mV.
    \textbf{c},~Magnetic field dependence of the value of $dI/dV$ at $V=-6$~mV.
  }
  \label{fig_Gd_spectrum}
\end{figure}%

We conducted spectroscopic-imaing scanning tunnelling microscopy (SI-STM) experiments on the Gd-terminated surface and found that the
$dI/dV$ spectrum and the spatial distribution of the electronic state exhibit
more complex behaviours than those of the Si-terminated surface.
Spatially averaged $dI/dV$ spectra are shown in Supplementary Figs.~\ref{fig_Gd_spectrum}(a)
and (b).
To clarify how the spectrum changes with magnetic field, the value of $dI/dV$ at
-6~mV as a function of applied magnetic field is plotted in
Supplementary Fig.~\ref{fig_Gd_spectrum}(c).
The $dI/dV$ value develops rapidly as increasing the magnetic field and
takes a kink at 1.4~T followed by a plateau. 
The plateau extends up to 2.2~T, which is the transition field between
\mbox{Phase\,I} and \mbox{I\hspace{-1pt}I} in the bulk. 
Then, $dI/dV$ at -6~mV shows a dip at 2.3~T, which corresponds to
\mbox{Phase\,I\hspace{-1pt}I}. 
In \mbox{Phase\,I\hspace{-1pt}I\hspace{-1pt}I}, right above the skyrmion phase, 
$dI/dV$ at -6~mV takes slightly higher value than that of the plateau and
continuously decreases with increasing the field. 
A step decrease is observed at the boundary between
\mbox{Phase\,I\hspace{-1pt}I\hspace{-1pt}I} and the FP phase.

The bulk magnetic phase boundaries manifest themselves as anomalies in the
field dependence of the $dI/dV$ spectrum of the Gd-terminated surface as well,
even though the appearance is different from that of the Si-terminated surface.
The major difference between the two terminations appears in \mbox{Phase\,I}.
The spectrum varies only slightly below 2.2~T for the Si-terminated surface.
In contrast, a clear kink at 1.4~T in Supplementary Fig.~\ref{fig_Gd_spectrum}(c) suggests
that the Gd-terminated surface possesses multiple magnetic states in the bulk \mbox{Phase\,I}.
This feature may originate from surface effects, because the magnetic Gd
layer is directly exposed to the vacuum for the Gd termination, whereas it is
sandwiched by RuSi layers for the Si-terminated surface.

\begin{figure}[t]
  \includegraphics[width=\columnwidth]{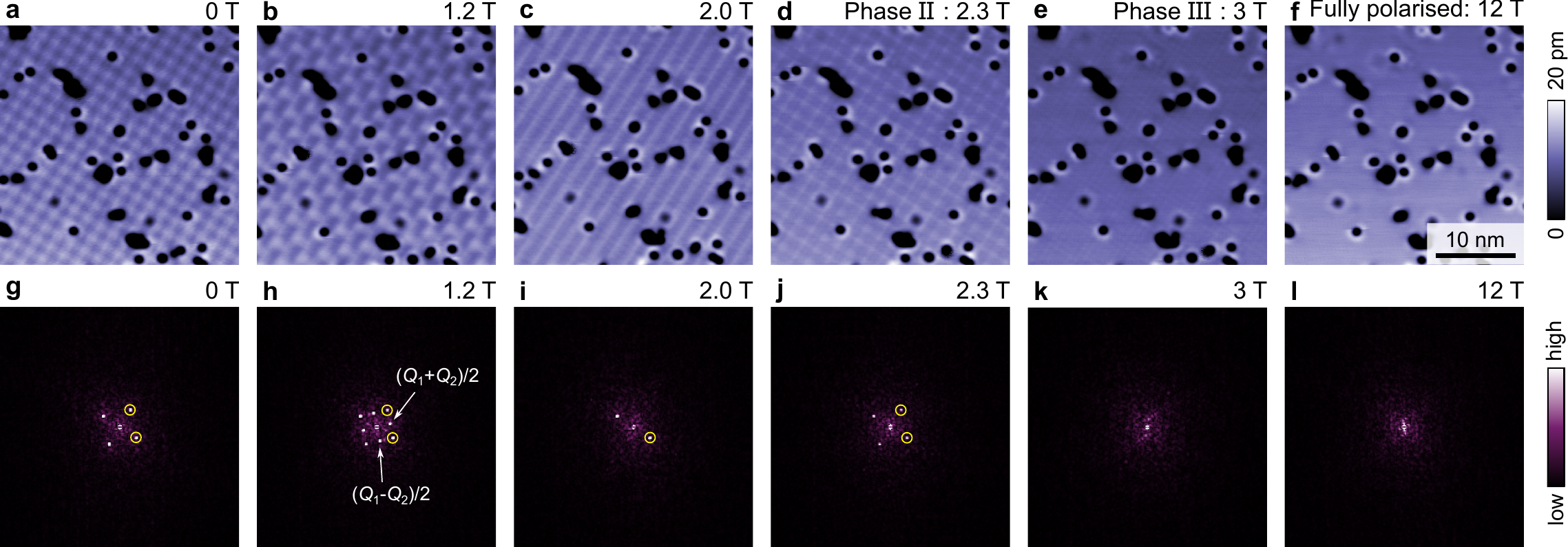}
  \caption{
    \textbf{a-f},~Constant-current topographic images of the Gd-terminated
    surface measured at 1.5~K.
    $V_{\mathrm{s}}=120$~mV and $I_{\mathrm{s}}=100$~pA.
    \textbf{g-l},~Fourier transforms of the filtered topographic images
    $z_\mathrm{filtered}$.
    Peaks marked with yellow circles correspond to $\mathbf{Q}_1$ or
    $\mathbf{Q}_2$ modulations.
  }
  \label{fig_Gd_all_phase}
\end{figure}%

We further investigated \mbox{Phase\,I} of the Gd-terminated surface by means
of constant-current topographic imaging and found three different periodic
structures, as shown in Supplementary Figs.~\ref{fig_Gd_all_phase}(a)-(c).
The constant-current topographic images represent the integrated local density of
states (LDOS) from the Fermi energy to the setup sample bias (0 to 120~meV for
these measurements) on top of surface corrugations.
Since atomic corrugations are not detected, the topographs here should be
governed by the spatial structures of the LDOS and indicate that the
Gd-terminated surface hosts three distinct states within bulk \mbox{Phase\,I}.
Topographic images for the other phases are also shown 
in Supplementary Figs.~\ref{fig_Gd_all_phase}(d)-(f) for comparison.

We found that deep defects in Supplementary Fig.~\ref{fig_Gd_all_phase}(a)-(f) dominate the
Fourier-transformed images and hide signals associated with the periodic structures.
Therefore, the topographic images are filtered using the following relation
before Fourier analysis.
\begin{equation}
  z_\mathrm{filtered}(\mathbf{r}) \equiv
  \begin{cases}
    z(\mathbf{r}) & \text{for } |z(\mathbf{r})-z_\mathrm{med}| \leq 0.3z_\mathrm{stdev} \\
    z_\mathrm{med} & \text{for } |z(\mathbf{r})-z_\mathrm{med}| > 0.3z_\mathrm{stdev},
  \end{cases}
\end{equation}
where $z_\mathrm{filtered}$ is the filtered height, $z$ is the measured height
in the topographic images, $z_\mathrm{med}$ is the median value of each topograph,
$z_\mathrm{stdev}$ is the standard deviation of each topograph, and
$\mathbf{r}$ is the lateral position.
Fourier transforms (FTs) of the filtered topographs are shown in
Supplementary Figs.~\ref{fig_Gd_all_phase}(g)-(l).
Modulation vectors are observed at $\mathbf{Q}_1$ and $\mathbf{Q}_2$ (yellow
circles), and $\frac{1}{2}(\mathbf{Q}_1 \pm \mathbf{Q}_2)$ (white arrows).
The $\mathbf{Q}_1$- and $\mathbf{Q}_2$-related modulation vectors suggest that
bulk magnetic structure plays an important role.
We speculate that additional surface effects may modify the magnetic structure
on the Gd-terminated surface, resulting in the multiple states within bulk \mbox{Phase\,I}.
Details are yet to be investigated.

\newpage
\section{A\lowercase{dditional data supporting the main observation}}

\begin{figure}[h]
  \includegraphics[width=0.9\columnwidth]{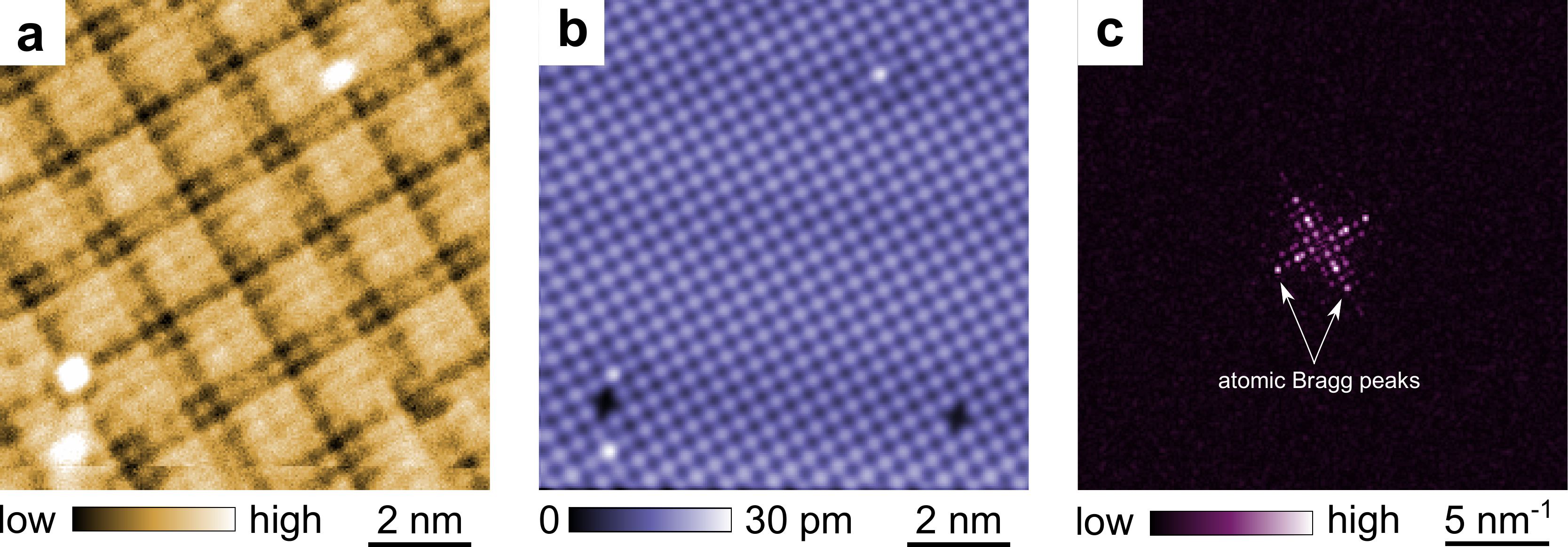}
  \caption{
    \textbf{a}, $L(\bm{r}, E=-20~\mathrm{meV}) = \frac{dI}{dV} / \frac{I}{V}$ map with
    higher spatial resolution measured at $\mu_0H = 2.3$~T and $T=1.5$~K
    (\mbox{Phase\,I\hspace{-1pt}I}, Skyrmion). 
    No additional fine features are observed.
    The scan was done from bottom to top.
    The discontinuity at the bottom of the  image shows that the skyrmion
    structure slides.
    Such motion is often observed soon after magnetic field was swept.
    \textbf{b}, Simultaneously measured topograph.
    Slides in atomic lattice was not observed.
    This suggests that the magnetic structures are not pinned to surface
    impurities.
    The setup sample bias voltage was $V_{\mathrm{s}}=100$~mV, tunnelling
    current was $I_{\mathrm{s}}=100$~pA, and the bias modulation amplitude was
    $V_{\mathrm{mod}}=5
    $~mV.
    \textbf{c}, Fourier transform of (a).
  }
  \label{fig_high_resolution}
\end{figure}%

\begin{figure}[h]
  \includegraphics[width=\columnwidth]{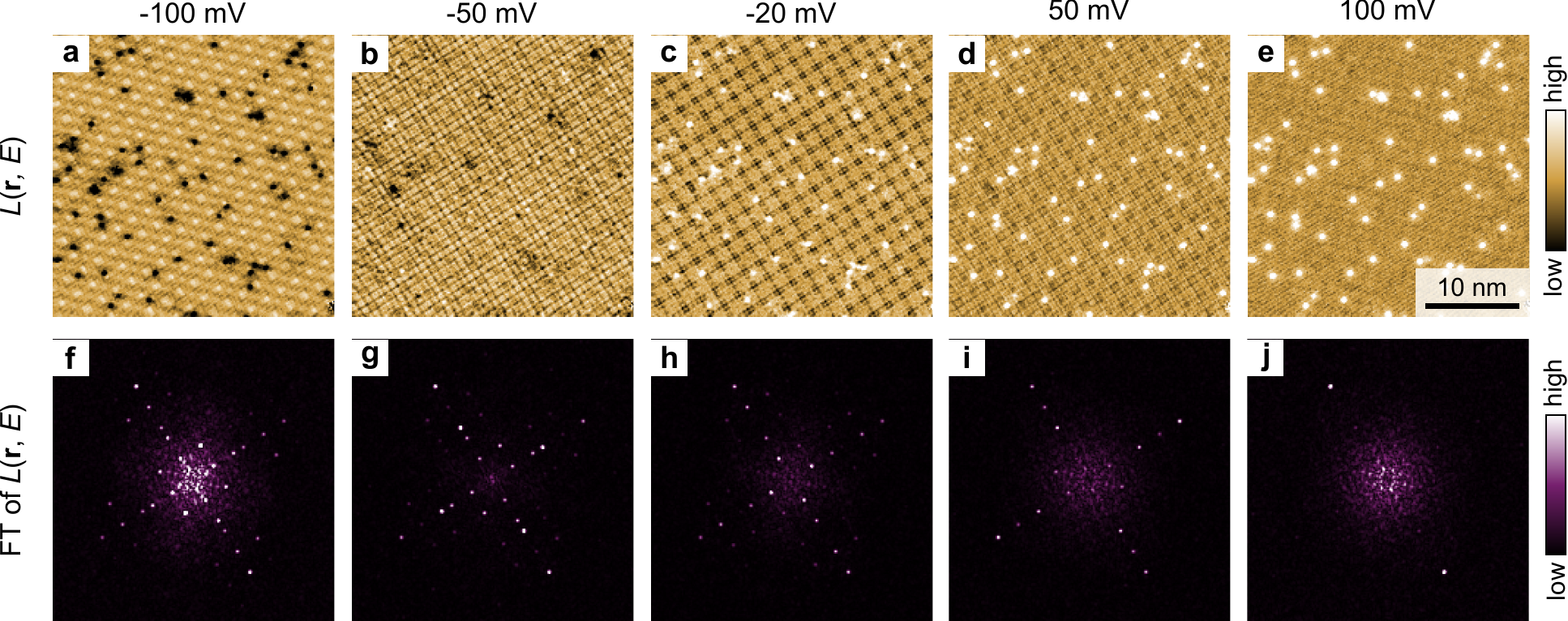}
  \caption{
    $L(\bm{r}, E) = \frac{dI}{dV} / \frac{I}{V}$ maps at different slices of
    bias voltages measured at $\mu_0H = 2.3$~T and $T=1.5$~K
    (\mbox{Phase\,I\hspace{-1pt}I}, Skyrmion).
    The position of thee Fourier spots do not depend on bias voltages while
    their intensity changes.
    $V_{\mathrm{s}} = 100$~mV, $I_{\mathrm{s}} = 100$~pA, and $V_{\mathrm{mod}}=5$~mV.
  }
  \label{fig_bias_dependence}
\end{figure}%

\begin{figure}[h]
  \includegraphics[width=0.8\columnwidth]{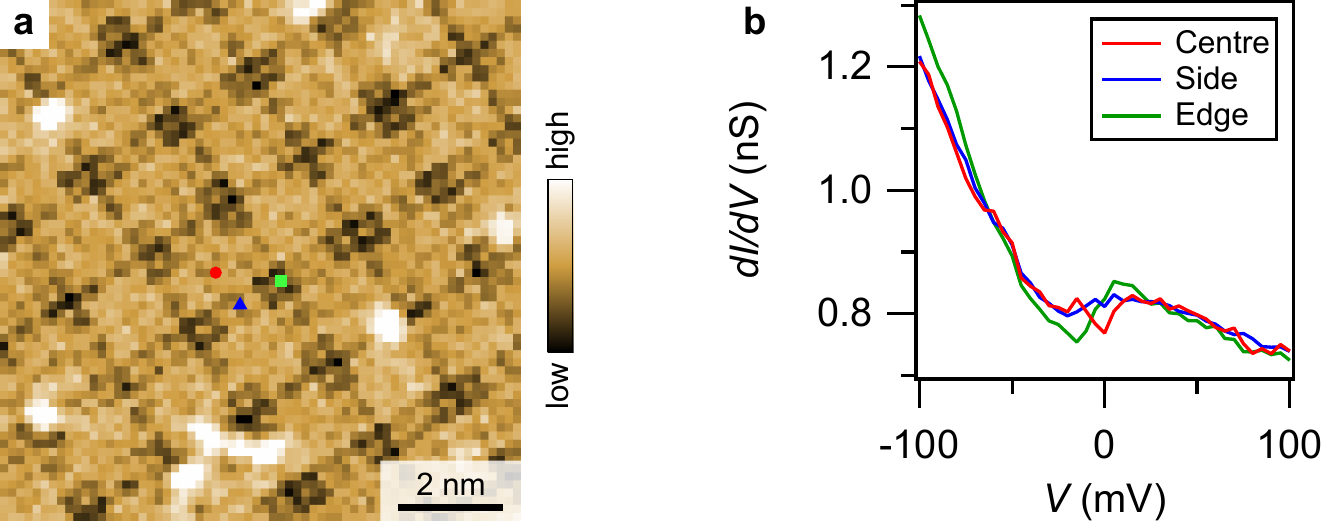}
  \caption{
    \textbf{a}, $L(\bm{r}, E=-20~\mathrm{meV})$ map measured at $\mu_0H = 2.3$~T
    and $T=1.5$~K (\mbox{Phase\,I\hspace{-1pt}I}, Skyrmion).
    Characteristic positions are marked with red circle (centre), blue triangle
    (side), and green square (edge).
    \textbf{b}, $dI/dV$ spectra at different potions.
    $V_{\mathrm{s}} = 100$~mV, $I_{\mathrm{s}} = 100$~pA, and $V_{\mathrm{mod}}=5$~mV.
  }
  \label{fig_point_spectrum}
\end{figure}%

\begin{figure}[h]
  \includegraphics[width=0.9\columnwidth]{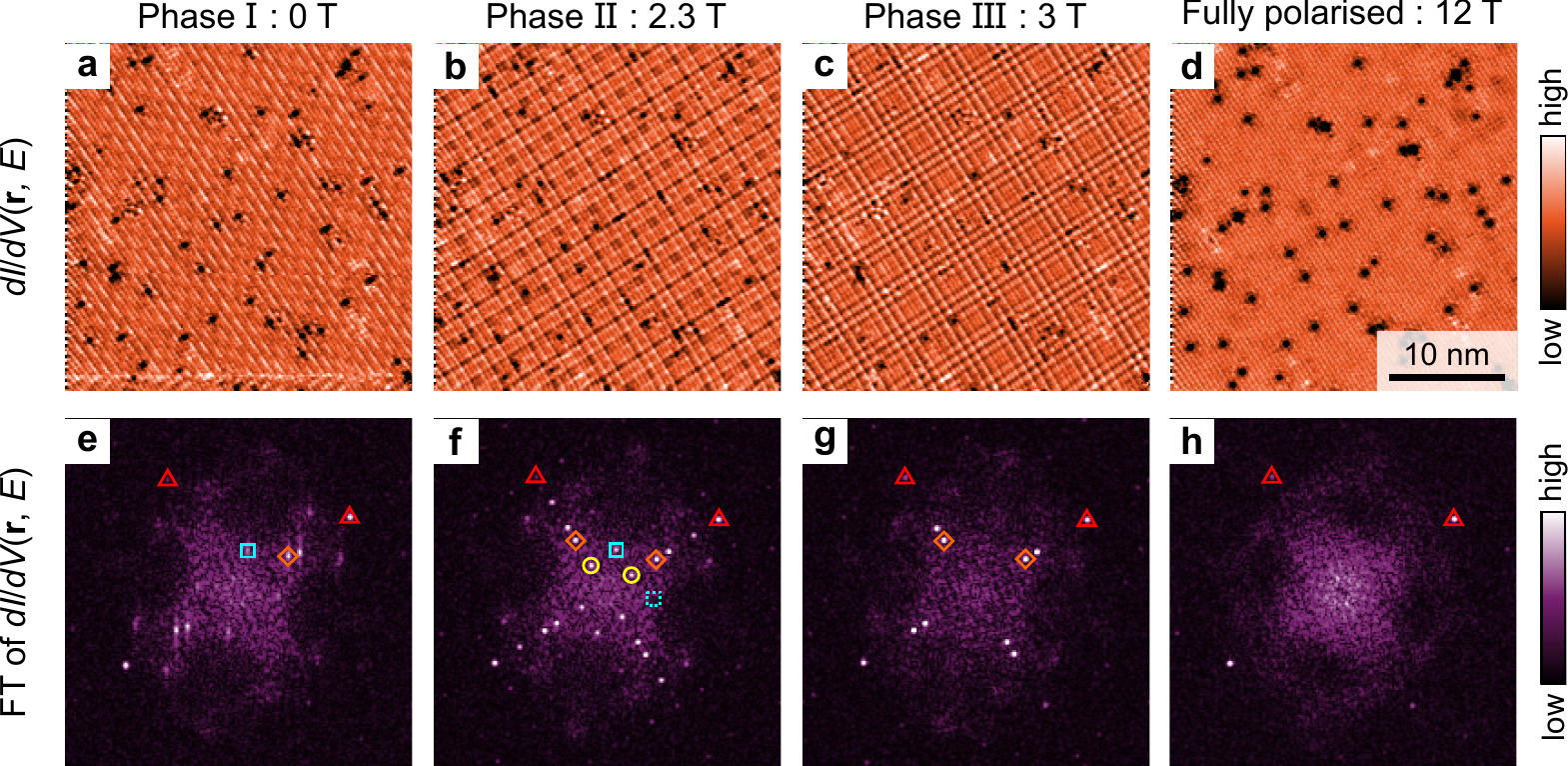}
  \caption{$dI/dV$ maps at $E=-20$~meV measured at $T=1.5$~K.
    These are the same sets of data as Fig.~3 in the main text but without normalisation.
    The normalisation does not affect to the Fourier $Q$-components.
    $V_{\mathrm{s}} = 100$~mV, $I_{\mathrm{s}} = 100$~pA, and $V_{\mathrm{mod}}=5$~mV.
  }
  \label{fig_field_dependence_no_normalize}
\end{figure}%

\begin{figure}[h]
  \includegraphics[width=\columnwidth]{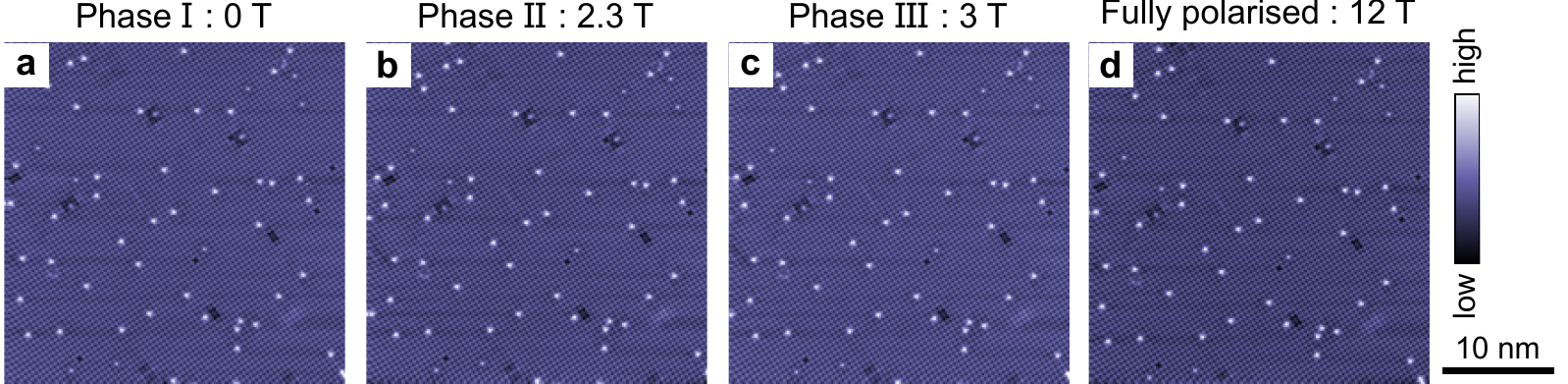}
  \caption{
    Topographic images simultaneously measured with Fig.~3a-d of the main text.
    The impurities as well as the scanning tip did not change during the
    measurement.
    $V_{\mathrm{s}} = 100$~mV and $I_{\mathrm{s}} = 100$~pA.
      }
  \label{fig_field_dependence_topo}
\end{figure}%

\begin{figure}[h]
  \includegraphics[width=0.5\columnwidth]{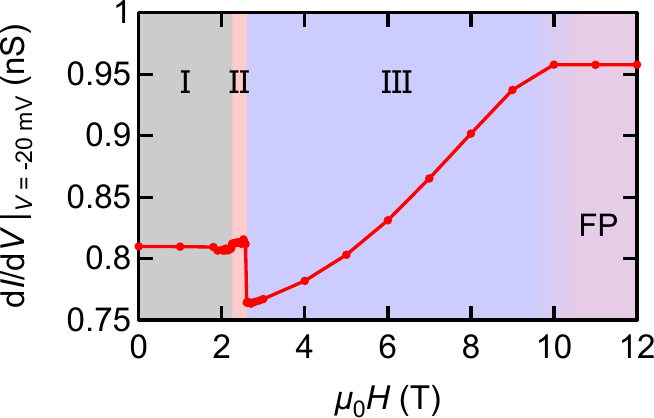}
  \caption{
    Magnetic field dependence of $dI/dV$ at $-20$~mV.
    First-order like transitions across \mbox{Phase\,I\hspace{-1pt}I} and 
    continuous change in \mbox{Phase\,I\hspace{-1pt}I\hspace{-1pt}I} are also
    seen at $V=-20$~mV.
    This behaviour does not depend on bias voltage as presented in Fig.~3k in the main text.
  }
  \label{fig_spectrum}
\end{figure}%

\begin{figure}[h]
  \includegraphics[width=0.5\columnwidth]{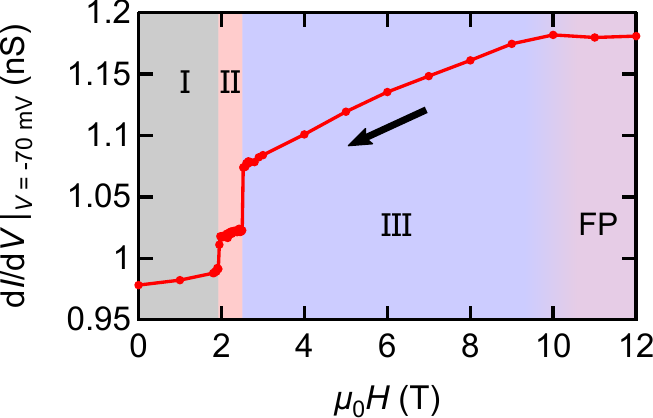}
  \caption{
    Magnetic field dependence of $dI/dV$ at $-70$~mV for magnetic field swept
    down from 12~T.
    The transition out from \mbox{Phase\,I\hspace{-1pt}I} is slightly shifted to
    lower value due to its first-order transition.
  }
  \label{fig_spectrum_down_sweep}
\end{figure}%

\begin{figure}[h]
  \includegraphics[width=0.9\columnwidth]{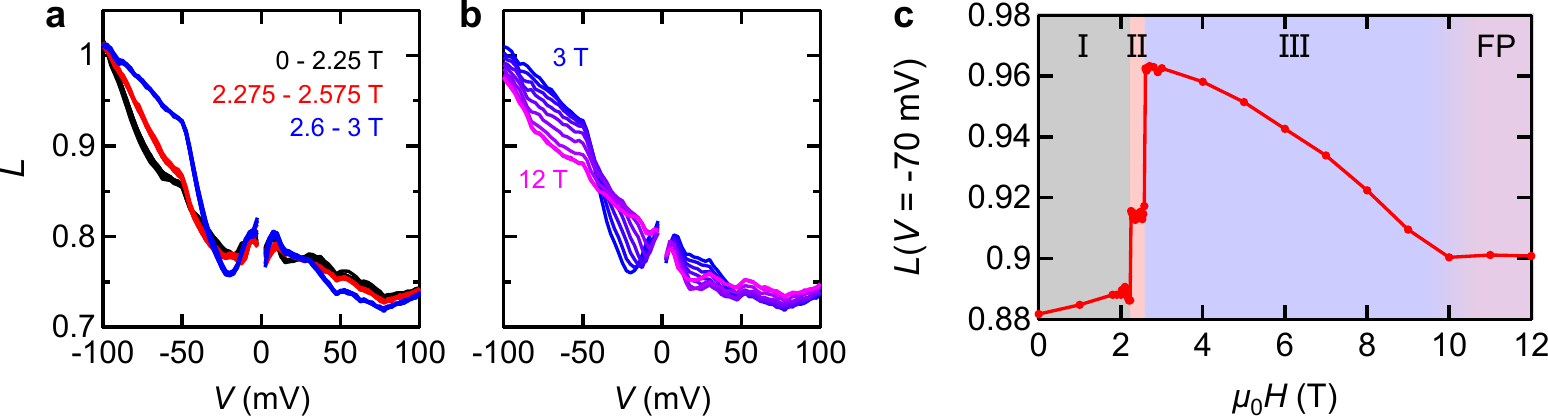}
  \caption{
    \textbf{i} and \textbf{j}, Spatially averaged normalised conductance
    $L=\frac{dI}{dV}/\frac{I}{V}$ curves \textbf{i} below 3 T and \textbf{j} above 3 T.
    $V_\mathrm{s}=100$~mV, $I_\mathrm{s}=100$~pA, and $V_\mathrm{mod}=2.5$~mV.
    \textbf{k}, Magnetic field dependence of the value of $L$ at $V=-70$~mV.
  }
  \label{fig_spectrum_normalize}
\end{figure}%

\begin{figure}[h]
  \includegraphics[width=0.7\columnwidth]{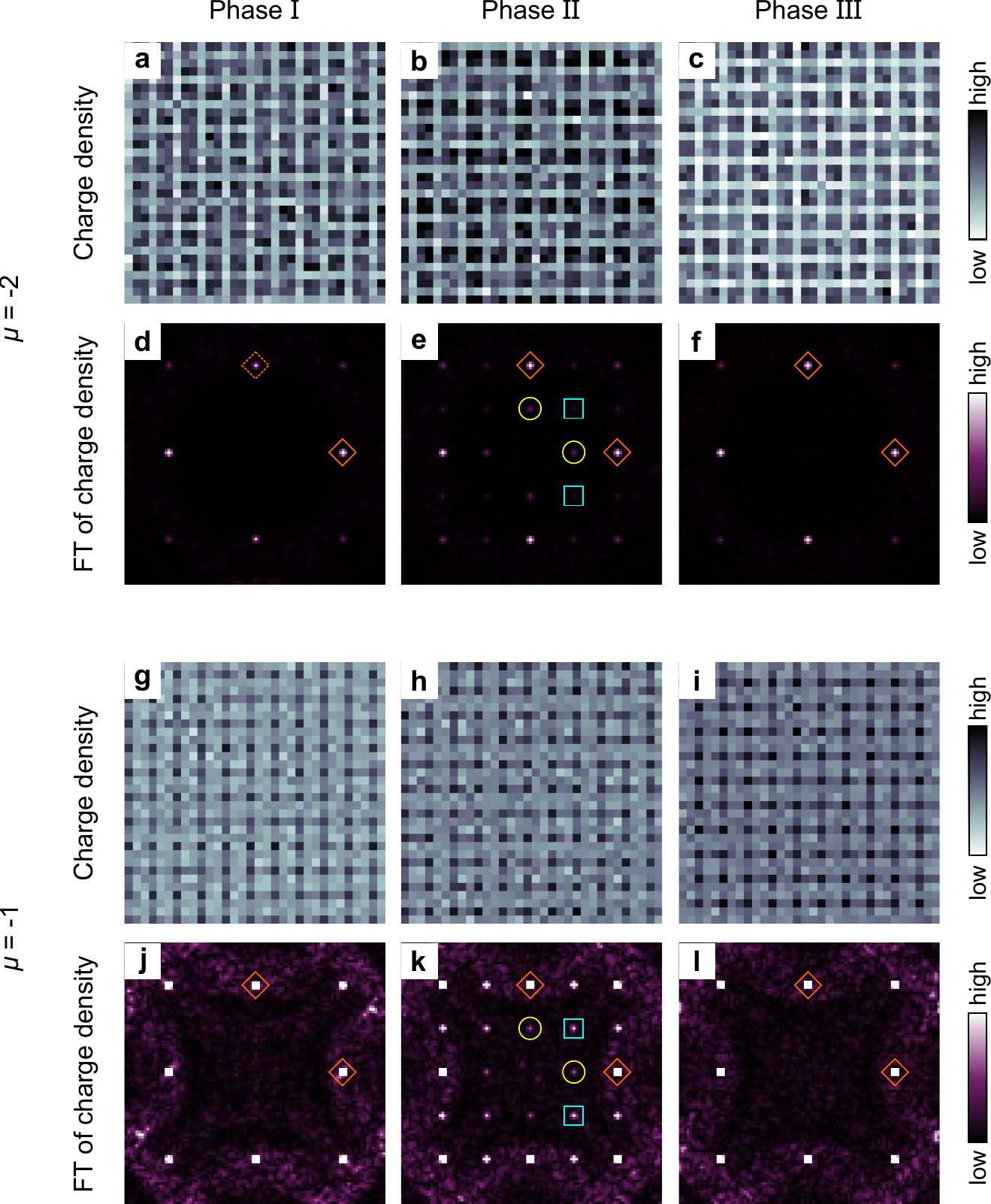}
  \caption{
    Charge density modulations calculated for different chemical potentials $\mu$.
    $\mu=-3$ is used for Fig.~4 in the main text.
    The Fourier $Q$-components do not depend on parameters we choose.
  }
  \label{fig_cdw}
\end{figure}%